# „Quantization of Reality" – another Epistemological Interpretation of Quantum Mechanics?

Dr. Carsten Reese, Hasenkamp 4, 28790 Schwanewede

Abstract

The epistemological interpretation of quantum mechanics is still in an unacceptable status. This becomes obvious if looking on the variety of interpretations currently under discussion. However, the physical community together with philosophers seem not to be willing to really face the problem and to find an explanation for the "oddness" of the foundation of nature, as the mathematical treatment of quantum physics can be judged as finished and perfect for the use. The epistemology of quantum physics is often treated as a negligible side effect – not worth to put effort in, as any outcome will not affect the calculation results at all. For the sake of human curiosity and to get a better understanding of the universe we are living in, this attitude should be changed. A possibly interesting new approach is outlined in this article, which may help to adjust the focus.

Introduction

For sure quantum mechanics is working. Quantum electrodynamics as a field theory is the most accurate physical theory at all. It describes the behaviour and the interactions between charged point particles and photons in a perfect manner (to the extend known today), e.g. the anomalous magnetic moment of the electron can be calculated and is in accordance with measurements down to the $10^{th}$ decimal [1]. Therefore we are not talking about correct calculations in quantum mechanics. We are talking about the mystery why it works, about the ontology of quantum physics. And here we have seen a lot of attempts, but unfortunately overall no convincing concept in the end. The very early "Copenhagen Interpretation", even criticised, is still in place, despite its obvious epistemological drawbacks and logical inconsequence. Copenhagen exactly stops explanation, where the questions start to hurt, and tells you to calculate instead of looking to the roots. This is not acceptable at all, as to be seen by the various attempts to unveil the secrets of quantum mechanics: hidden variables, decoherence, pilot waves, many world interpretation etc. We will see later on.

Basic Considerations

Locality and/or realism – at least one of these cornerstones of classical physics and common sense has to be abandoned. This has been proofed by experiments on the EPR-paradox [2,3] together with the Bell inequality [4]. Quantum mechanics is at least non-local, but in a way which does not violate causality. The instantaneous collapse of the wave function of special separated, but entangled quantum objects does not include any information transport, if locality is dismissed and/or the wave function is deemed not to have any reality, but only the measurement itself.

To resume the process of entanglement very shortly: a quantum system of two objects is prepared in a way that is can be described by one common wave function only. This is achieved by coupling e.g. the spin parameter of object A to the spin parameter of object B in a way that, if object A shows "spin up" in a measurement, object B necessarily shows "spin down", and vice versa, by using conservation principles for the spin. If the objects are separated after the preparation, the entanglement still exists, in fact we are talking about one quantum object only. As soon as one of the spin parameters is measured, the spin of the other "part" of the object is fixed – without any delay. This is what Einstein called the "spooky action at a distance" ("spukhafte Fernwirkung"), which he wanted to logically exclude by constructing the EPR paradox, but has been shown to exist by experiments. By proving the violation of Bell´s inequality it has been shown that at the time of the entanglement no information is available, from which the result of the measurement could be deducted, therefore no "hidden variables" exist. The manifestation of the result is indeed only at the time of the measurement.

In a classical view, the information from object A to object B would have to be transferred with infinite velocity, explicitly faster than light. But in the view of quantum mechanics there is only one object, so no need to transfer anything, even if the parts are separated in space. Therefore quantum mechanics is necessarily a non-local theory.

What is it about with the time between entanglement (the last interaction) and the measurement? Within this time frame there is definitively no information about the state of the spatial separated, entangled quantum object available. All what is known is the possibility for different states, which obviously exist in parallel. This is against our understanding of the terminus "reality". It leads to the conclusion that the wave function does not have any real analogy.

If the wave function is without any real analogy in our world, it is wrong to talk about a superposition of states (a mixture of spin up and spin down). In fact we do have only abstract possibilities. The wave function is a mathematical construct only to calculate the probability of the realisation of these possibilities at a certain time, if decoherence occurs by performing the measurement or, more generic, allowing the information to be available.

"Reality" in the sense we are using this term is therefore limited to a series of decoherences with intermediate phases of non-reality with possibilities, how reality could develop. Reality is quantized, because in the intermediate phases no information about the system state is existing (anywhere in the universe). Applying logic consequently leads to the conclusion that the system cannot be real in this phase. If I try to get the information, I need to do a measurement, implying the "collapse of the wave function" or decoherence, so a new "reality point" would be generated. If there is no way to get information we cannot assign any "reality" to the quantum system state, but only "possibility".

According to the decoherence theory [5, 6] the sequence of decoherences for macroscopic systems is very dense, because of permanent interactions with the environment decoherence occurs within very short time scales. Our impression of the environment therefore is a permanent presence of reality. If we avoid any decoherence, which is possible in the preparation meanwhile also with larger quantum objects, the non-real phases are getting longer and become visible in the way, that we don´t have information, but only possibilities for the development of reality. The mathematical description of these effects is what we call quantum mechanics.

Einstein said "God does not play dice" [7], but this needs to be changed to "God does not know everything" in this picture, as information about a quantum systems is not available during coherent phases by principle.

Quantized Reality in the Context of Existing Interpretations

The context of the presented interpretation of quantum mechanics can be seen towards existing interpretations in the following way:

The Copenhagen interpretation in its fundamental and unquestionable elements is acknowledged as valid, of course. This means, quantum mechanics is logically accepted as a non-local and non-real theory. However, the Copenhagen interpretation in its different modifications stays unacceptable vague and contradictory concerning the measurement problem (this is meant also in view of the transition quantum mechanics to classical physics) and the role of the observer. The quantized reality approach tries to overcome at least some of these problems by including two other known approaches: the decoherence theory and information based concepts for the interpretation of quantum mechanics.

The decoherence theory in the form as proposed by Zeh is not in line with the Copenhagen interpretation, as the change of a quantum state is seen as a continuous process. This means, the collapse of the wave function has a certain duration, the so-called decoherence time. This approach seems to be problematic, because it defines decoherence and collapse of the wave function as the same process. Especially if talking about long decoherence times as for well isolated quantum systems, the contradiction looks obvious. However, an important aspect is the consideration of interaction with the environment by giving decoherence times. It shows, that a quantum mechanical description of macroscopic objects is correct, but just meaningless, as the wave function of the system would collapse and have to be changed permanently due to the inclusion of more and more interacting particles. The approach therefore allows us to get a better understanding about the transition of quantum mechanics to classical physics.

Information based interpretations of quantum mechanics [8] are also giving an important contribution. Information is treated as a physical unit, and is therefore expected to have real influence on the physical reality, which means, information has an impact on measurement results. If a specific information is available, the measurement result will be limited in its variety. If the information is not available (cannot be made available for principle reasons), there are more possible results for a measurement – independent from the physical state of the system. This is in accordance with the Copenhagen interpretation, even if the ontological definition is not using this element explicitly.

The difference between the presence and the lack of information leads to physical measurable effects. This concept is less surprising than one might think. E.g. electrons are indistinguishable. Just because of this fact (there is no information available to distinguish one electron from another *by principle*) and that the wave function is antisymmetric, the Fermi-statistic applies. Same is for the symmetric wave functions of the Bosons and the Bose-Einstein statistics. If there would be any possibility *in principle* to distinguish the identical particles, in both cases these statistics would not describe nature. In other words, the validity of these statistics is a proof that the particles are indistinguishable.

Using this insight in the other direction, the non-real ("coherent") phases, which have to be treated as statistical and calculable possibilities, give proof that no more information is available in principle, which is the same as concluding that no hidden variables are present.

Following this epistemological understanding, some of the classical quantum mechanics problems can at least partially be looked at in a different light. This is done in the following chapters.

It may be worth to note here that it seems a very similar view, however not given in detail and not treated as an individual interpretation, has been described by Anton Zeilinger [9].

The Collapse of the Wave Function

A discontinuous change of the wave function occurs in the moment of an interaction, as the quantum object is now described by a wave function of the former quantum object and the interacting quantum object. This can be an interaction aiming on getting a measurement result, but also an "unintentional" interaction with quantum objects of the environment. The interaction results in information gain on the system – a measurement result, independent whether there is an observer or not, whether there is a measurement device or not. Within this interaction necessarily a part of the possibilities existing before (e.g. probability densities as a solution of the Schrödinger equation of the quantum object) are reduced from a probability ("could be") to a definite state ("is"). In a different view we can say: a point in reality became manifest, where before just possibilities existed. The coupled, new system now evolves according to the Schrödinger equation and stays non-real: again we are just talking about possibilities and probabilities up to the point of the next interaction and therefore the next reality point.

Also by this view it does not become clear why a certain measurement result becomes manifest. It stays open how a "selection" is made out of the possibilities to become reality. If this specific issue is understood as the "measurement problem", there is unfortunately no obvious progress due to this approach.

In the Copenhagen interpretation the process of a "measurement" refers to a macroscopic measurement device. To my understanding this is not acceptable as it uses a physically not allowed anthropocentric world view, which also generates big problems in the differentiation between classical and quantum mechanics terms and definitions. Within the interpretation given in this article a measurement is generally referred to as the availability of information in principle, not information for a measurement device or any observer. In this sense also any problems generated in the frame of consciousness like "Is there a moon, if nobody is looking at it?" are without any meaning. Also the question "what is a measurement?" has a very clear answer: a measurement is the occurrence of the collapse of the wave function by an interaction, which generates information gain.

Therefore, a measurement results in an availability of information about a quantum object in principle. It does not matter if the information is processed by a measurement device or consciousness.

Side Note: it is very surprising how the Copenhagen interpretation, which took the logical consequences of quantum mechanics calculations very serious, is dealing with the question concerning measurement and observer. It obviously uses anthropocentric and contradictory formulations. Even more surprising, that this has not be corrected over a timeline of about 90 years, despite the fact, that the problem was well known as the Heisenberg cut and has been discussed in length.

Schrödinger´s Cat

The cat is interacting with the environment. As a cat is a macroscopic object, the sequence of reality points is very dense (about $10^{26}$ per second, as estimated by the coherence theory [10]). As long as the radioactive atom is not decayed, the status "cat is living" manifests every $10^{-26}$s therefore. Within two of the interactions with the environment the cat is dead and alive, both states are possibilities because no information about is available. Once the decay takes place, it takes again about $10^{-26}$s to have the status "cat is dead" manifested by decoherence induced collapse of the wave function. This does not require a "conscious observer" to open the chest and verify what happened.

Therefore this approach of quantized reality does not generate a problem concerning Schrödinger´s paradox (in fact, in this view there is no paradox at all), as well as with the very similar problem of indifferent indications of pointers during the measurement of superpositions.

Many-world-interpretations

Basically the quantized reality interpretation does not give a hint pro or con of many-world-interpretations, as it is not proposing any mechanism for a decision on one or the other possibility. As long as this "measurement problem" is kept open, many-world interpretations are always a possible interpretation. However, there is at least one clear statement following from quantized reality, and this is concerning the number of parallel worlds. If the many-world-interpretation is correct, this number must be incredible high. Taking the Copenhagen interpretation, one could limit the number of parallel worlds, if it is said, that a split only occurs in case of a quantum measurement with a certain result – at least this would be one possible interpretation. In this case we would have to deal with a number of universes which seems to be "manageable". To the opposite, quantized reality requires permanent decisions and splits, which would lead to parallel universes in each case. Also in these parallel universes immediately interactions would lead to new parallel universes, and so on. The decimal powers occurring here would not be presentable after just one second – literally, as we would have decimal powers in decimal powers in decimal powers…

The question may be allowed, if such an interpretation, which in addition is obviously not verifiable (or better not falsifiable), does make any sense at all. But, as said before, there is no stringent logical argument following from the quantized reality against many worlds.

Bohm´s Mechanics

Bohm´s mechanics tries to preserve the idea of reality for the wave function, respectively for the coherent phases. It is therefore incompatible with the approach as presented here, in fact it is contradictory. In case, however this might become possible, the idea of quantized reality will be proofed, the ontology of Bohm´s mechanics would be falsified – and the other way round.

But looking purely at the mathematical approach of Bohm´s mechanics it is not just to be neglected, also in view of quantized reality. During non-real phases position and momentum of a "particle" are not known precisely, and the imprecision increases with time. In the view of quantized reality it is proposed not to assign the idea of real existence of a quantum object during these times, because no information is available by principle. However, the mathematical description of the development of the uncertainty can be done in different ways, either by a wave function, or by using a particle and a pilot wave, or by using Heisenberg's matrix mechanics.

Particle-Wave Dualism

In fact the quantized reality approach elegantly solves this problem of quantum mechanics. The problem in principle is, that photons or other "particles" show a wave character (interferences) in certain experiments, but clearly particle character when detected. A closed model to understand this dualism, that means a merge of the description of wave and particle in one model, seems to be unreachable because of the contradictions of particle and wave character. Quantum field theories do not show this problem at all, because interactions and particles are treated as discrete excitations of fields. However, this does not solve the descriptive contradiction of the dualism.

In this approach wave and particle character are naturally separated *in time*. The wave character describes the non-real coherent phase. As an example, behind a 50% beam splitter the possibility of two ways exist. As no information is available which path is taken, a description in the particle view is impossible. Obviously it is necessary to find a different description during this non real phase. It has been show experimentally that a description based on a wave character is adequate. But as we do not assign any real manifestation for this phase, we cannot talk about a characteristic of a quantum object, but just give a mathematically correct calculation to derive possibilities. However, it is surprising that obviously a description in the wave form, which leads to effects like interference phenomena, is the correct way to calculate these non-real phases. It would be interesting to evaluate, if this is a logical consequence out of the treatment as possibilities or if other descriptions, not leading to interferences, could be constructed as well. In the latter case one would have to answer the question why specifically the wave description is successful. In other words, it should be shown that based on quantized reality a necessary outcome would be the De-Broglie equation.

The detection of a photon, electron or any other quantum object always takes place via an interaction as a localisation. If a reality point is generated, we "see" a particle, and this is the classical expectation. "Real" therefore is the particle description, whereas the wave description builds the mathematical basis for the calculation of possibilities during two decoherence points. A simultaneous view respectively a closed model to cover both aspects of wave and particle is therefore not necessary.

Delayed Choice Experiments

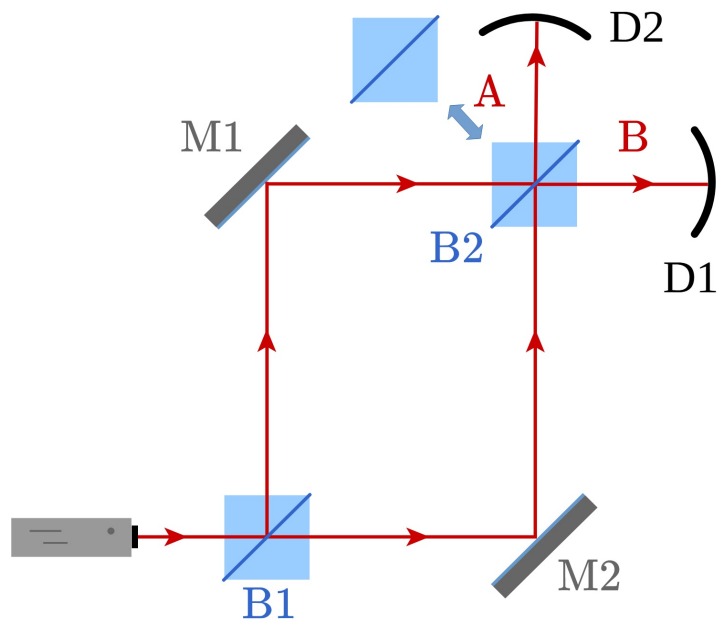

Figure 1: Mach-Zehnder Interferometer with delayed choice principle (source: https://de.universaldenker.org/illustrationen/610, modified for delayed choice)

Looking at the classical Mach-Zehnder interferometer setup (Figure 1) with a single photon source, a 50% beam splitter B1, two 100% reflective mirrors M1 and M2, and two photon detectors D1 and D2 as well as the second beam splitter B2, which is placed or not only after the photon has passed the first one (for the delayed choice), we find interference effects only, but always if the second beam splitter is inserted. Without second beam splitter we get a 50 to 50 percent count for the photons, with second beam splitter we can adjust the counts from 0 to 100 (for detector 1 in comparison to detector 2) to 100 to 0 as an interference effect by carefully adjusting the two different path lengths.

It is argued, that the photon needs to "decide" to take one or the other path (as a particle) as long as the second beam splitter is not present, because we don´t see interference effects in the detectors if summing up lots of single measurements. In case the beam splitter is present, the photon takes both paths at a time, actually as a wave. If it would be like that, some problems arise if the second beam splitter is inserted after the photon has passed the first one, as the decision about particle or wave has been made already (delayed choice). Experimental results show that always the final state of the experimental setup determines the outcome, as expected by quantum mechanics, but in principle forcing the photon to make its decision after it already passed the decision point. On the other hand, the photon would need to know about the second beam splitter, which is an odd argumentation as well. So, in the end quantum mechanics shows, that measurement results are those, which could just be assigned to something which is in classical physics a wave in one case or a particle in the other.

Again, in the view of quantized reality the problem does not exist in this way. After the first beam splitter, there is no information about which path is taken. The position information is generated only at the moment of the detection in the photon counter, which is either the first or the second by a 50 to 50 chance, or with a distribution according to the interference terms in case the second beam splitter is in. Any question about which path has been taken is meaningless, as it can never be determined by principle. "It must have come this way" in case of just one beam splitter is a meaningless statement, because there is absolutely no way to verify it, which means, to get the information that this really happened. There has been a probability for one way or the other during the transit time, and this non-real state cannot be changed by a logical deduction reaching back in time. There has been a probability for different paths, and even if a detector clicks, it does not mean that the photon has taken a certain path and therefore behaved as a particle. It started in at the photon source and ended up in detector D1 or D2, that is all about it. In view of the quantized reality there is no difference concerning the behaviour of a photon between the experiment with or without second beam splitter, and also not, if this is run as a delayed choice.

A very good overview of delayed choice experiments and also the interpretation of the results is given in [11]. I would like to add a citation here:

"It is a general feature of delayed-choice experiments that quantum effects can mimic an influence of future actions on past events. However, there never emerges any paradox if the quantum state is viewed only as a "catalog of our knowledge" [12] without any underlying hidden-variable description. Then the state is a probability list for all possible measurement outcomes and not a real physical object."

This appears to be in line with the concept of quantized reality, but misses the final step towards accepting the brutal logic behind this – it is not valid for delayed choice only, it is a general concept.

## Conclusion

The presented concept of quantized reality can be a very first step only. It hopefully shows that we still have clearly a lack of understanding of the foundations of quantum mechanics, but that it will not be impossible to go a step further, even if trials have been made since nearly 100 years. If a better understanding on the foundations of quantum mechanics is possible, this may lead to new insights as well.

Some important open questions to further develop the concept of quantized reality:

- Can it be shown that, based on Heisenbergs uncertainty relation and the concept of possibilities starting after a reality point only, that a mathematical description of a quantum system is *necessarily* leading to the Schrödinger equation or an equivalent mathematical description? In other words, the concept of quantized reality, which has been presented in the form of arguments here, needs a mathematical basis and description.

- Can the process of interaction be described in more detail? When does an interaction occur? As quantum physics is non-local, interactions may not be limited to quantum objects sharing a certain region in space. Does it make sense in the frame of the concept of quantized reality to ask these questions at all?

- Based on the concept, can we find any ideas towards a solution of the measurement problem (understood as the *process of decision* for a certain reality)? As we have seen above, an interesting point maybe that the tool of logical deduction to a former point in time ("it must have taken this path") is not allowed. The past with its variety of possibilities does not change after the collapse of the wave function.

- How should time-independent solutions of the Schrödinger equation (e.g. the ground state of a hydrogen atom) be treated in view of the concept? Such states would imply a permanent reality, as there are no possibilities, but a certain state with a valid information.

In doing these next steps, one may start to unveil the "great smoky dragon" of John Archibald Wheeler.

References:

[1] Hanneke, D.; Fogwell Hoogerheide, S.; Gabrielse, G. (2011). "Cavity Control of a Single-Electron Quantum Cyclotron: Measuring the Electron Magnetic Moment" (PDF). Physical Review A. 83 (5): 052122. arXiv:1009.4831. Bibcode:2011PhRvA..83e2122H. doi:10.1103/PhysRevA.83.052122

[2] A. Einstein, B. Podolsky, N. Rosen: Can quantum-mechanical description of physical reality be considered complete?, Phys. Rev. 47 (1935), S. 777–780 doi:10.1103/PhysRev.47.777

[3] Alain Aspect, Philippe Grangier, and Gérard Roger: Experimental Realization of Einstein-Podolsky-Rosen-Bohm Gedankenexperiment: A New Violation of Bell's Inequalities. In: Phys. Rev. Lett. Band 49, 1982, S. 91–94, doi:10.1103/PhysRevLett.49.91

[4] Bell, J. S. (1964). "On the Einstein Podolsky Rosen Paradox". Physics Physique Физика. 1 (3): 195–200. doi:10.1103/PhysicsPhysiqueFizika.1.195

[5] H. D. Zeh: On the Interpretation of Measurement in Quantum Theory. In: Foundations of Physics. Band 1, Nr. 1. Springer Verlag, 1970, S. 69–76

[6] Joos, E., Zeh, H.D.: The emergence of classical properties through interaction with the environment. Z. Physik B - Condensed Matter 59, 223–243 (1985). https://doi.org/10.1007/BF01725541

[7] Albert Einstein, Hedwig und Max Born: Briefwechsel 1916–1955. Rowohlt Taschenbuchverlag, Reinbek bei Hamburg, 1972, S. 97f.

Einstein in einem auf den 4. Dezember 1926 datierten Brief an Max Born

[8] z.B. A. Grinbaum, Reconstruction of Quantum Theory, Brit. J. Phil. Sci. 8 (2007), S. 387–408.

[9] Zeilinger, Anton, Einsteins Schleier. Die neue Welt der Quantenphysik. München 2003, ISBN=3-406-50281-4; S.194

[10] E. Joos et al.: Decoherence and the Appearance of a Classical World in Quantum Theory, Springer 2003, ISBN 3-540-00390-8

[11] Xiao-song Ma, Johannes Kofler, Anton Zeilinger: Delayed-choice gedanken experiments and their realizations, REVIEWS OF MODERN PHYSICS, VOLUME 88, JANUARY–MARCH 2016

[12] Schrödinger, E., 1935, Naturwissenschaften 23, 844.